\documentclass{Interspeech}

\interspeechcameraready

\title{SimClass: A Classroom Speech Dataset Generated via Game Engine Simulation For Automatic Speech Recognition Research}

\author[affiliation={1}]{Ahmed Adel}{Attia}
\author[affiliation={2}]{Jing}{Liu}
\author[affiliation={1}]{Carol }{Espy-Wilson}

\affiliation{Electrical and Computer Engineering}{University of Maryland}{}
\affiliation{College of Education}{University of Maryland}{}
\email{aadel@umd.edu, jliu28@umd.edu, espy@umd.edu}
\keywords{speech recognition, dataset}

\usepackage{comment}

\begin{document}

\maketitle

\begin{abstract}
The scarcity of large-scale classroom speech data has hindered the development of AI-driven speech models for education. Public classroom datasets remain limited, and the lack of a dedicated classroom noise corpus prevents the use of standard data augmentation techniques. 

In this paper, we introduce a scalable methodology for synthesizing classroom noise using game engines, a framework that extends to other domains. Using this methodology, we present SimClass, a dataset that includes both a synthesized classroom noise corpus and a simulated classroom speech dataset. The speech data is generated by pairing a public children’s speech corpus with YouTube lecture videos to approximate real classroom interactions in clean conditions. Our experiments on clean and noisy speech demonstrate that SimClass closely approximates real classroom speech, making it a valuable resource for developing robust speech recognition and enhancement models.
\end{abstract}

\section{Introduction}

The performance of AI models is defined by its training data, and the model's ability to learn from this data. Speech models are no exception. For instance, Whisper \cite{radford2023robust} was able to achieve state-of-the-art performance in the Automatic Speech Recognition (ASR) task by training on more than half a million hours of transcribed speech scrapped from the internet. 

Naturally, this scalability does not apply to low-resource settings. For the same ASR task, low-resource languages lag behind English and other high-resource languages precisely due to the lack of high-quality labeled data for these languages \cite{san2024predicting, nowakowski2023adapting}. Beyond linguistic variability, even some English language tasks suffer from data scarcity. Challenging acoustic conditions like high background noise and multi-speaker environments usually require more training data for robust performance. However, for some domains, like classroom speech and children's speech, there is a severe data scarcity, since children are considered a protected class and speech is considered Personally Identifiable Information (PII) \cite{coppa2024}.  As a consequence, releasing public datasets including children's speech is complicated.

Prior research into developing classroom and children ASR tools identified data scarcity as a major hurdle to improving performance \cite{attia2024cpt, southwell2024automatic, southwell2022challenges, fan2022towards, fan2024benchmarking}.   

While there are some children ASR datasets that are public,\cite{pradhan2023my, shobaki2000ogi, batliner2005pf_star},  most classroom corpora are not. As a result different researchers working on the same problem often work with different datasets. Getting access to protected classroom speech corpora is a complicated task, so each research group tends to collect their own data, but cannot publish it \cite{attia2024cpt, southwell2022challenges}. This lack of dataset exchange negatively affects the progress of research and reproducibility of research.

Some prior research has investigated data augmentation methods that adjust the acoustic properties of adult speech to match that of children, such as pitch perturbation \cite{fan2024benchmarking}. These methods do not apply to classroom speech. The main defining challenge of classroom speech is the existence of a high level of children's babble noise - the type of noise resulting from multiple background speakers overlayed with speech from the target speaker. While adult babble noise corpora exist \cite{snyder2015musan, font2013freesound}, no such corpus exists for children's babble noise.

In this paper, we present our efforts towards solving both issues, the scarcity of classroom data, and the lack of children's babble noise corpora. We create a clean classroom noise corpus, by combining the My Science Tutor (MyST) corpus with instructional adult speech from lectures from MIT OpenCourseWare (MIT - OCW) and Khan Academy. These channels' videos are licensed under Creative Commons BY-NC-SA \cite{cc-by-nc-sa}, which permits non-commercial use, adaptation, and redistribution with proper attribution, provided that any derivative works are also shared under the same license. However, since classroom speech is noisy by nature, we synthesize classroom noise, primarily children's babble, through the Unity Game Engine. The Unity Game Engine provides high-fidelity audio simulation in 3D virtual space, which accurately simulates the acoustic characteristics of the classroom environment, as well as the spatial characteristics of different audio sources.

We call our \textbf{Sim}ulated \textbf{Class}room dataset \textbf{SimClass}. We plan to make SimClass publicly available at no cost to researchers, by the time of the camera-ready manuscript. SimClass is the largest and only public classroom speech dataset. In addition, the existence of a clean and noisy version of the dataset, which is not possible with real classroom recordings with naturally occurring noise - opens the door to many tasks that were not possible, such as speech enhancement. SimClass mainly simulates elementary school STEM classes as a result of its constituting data, but our methodology outlined in this paper paves the way for simulating other kinds of classrooms given different kinds of data.

\section{Datasets}
\label{sec: data}
\subsection{Datasets Used To Create SimClass}
\subsubsection{My Science Tutor (MyST)}
The MyST corpus is the largest publicly available children's speech corpus. It consists of 393 hours of conversational children's speech, recorded from virtual tutoring sessions in physics, geography, biology. The corpus spans 1,371 third, fourth, and fifth-grade students. Around 210 hours of the corpus were accurately transcribed after filtering out weak, inaccurate transcriptions following the method outlined in prior work \cite{attia2023kid}. We utilize the transcribed portions of the children's speech for our clean speech base and use the speech from the untranscribed portion for noise simulation. We have verified that there is no overlap in speakers between the transcribed and untranscribed portion, to avoid having the target speech from the primary child speaker also be present in the background.
\subsubsection{Khan Academy}
Khan Academy is a popular YouTube channel hosting thousands of educational videos targeted at different school-age grades covering a variety of subjects and topics. Although Khan Academy videos provide a good match in both topic and target audience, as there is a variety of elementary STEM playlists, they mainly have the same single speaker, which can limit the diversity of the data. Additionally, these videos are recorded virtually on a tablet, not in a classroom which can affect the acoustic properties of the lecturer. However, to benefit from the agreement on the subject matter and topics, we include videos from this channel in our dataset. 

We chose 7 playlists discussing topics in second and third-grade mathematics. The total duration of the videos in these playlists is about 7.5 hours, and they include high-quality audio accompanied by gold-standard human transcripts. We will provide a full list of videos used on our GitHub repository in the camera-ready manuscript.

\subsubsection{MIT OpenCourseWare (OCW)}
MIT OCW \cite{MITOCW_YouTube} is an online repository of recorded lectures from 2,500 MIT courses. Although the classes in OCW are college-level and graduate-level classes, they still provide a useful resource for elementary school STEM classes, as the acoustic properties of instructional speech are evident in the videos. However, to help bridge the gap between the linguistic components of college-level courses and elementary school STEM classes, we picked videos from 10 courses in calculus and linear algebra, totaling around 174 hours. 

These videos include high-quality clear audio from a variety of lecturers with different accents, accompanied by high-quality human transcriptions, verified to be uploaded by the channel and not generated by YouTube's speech-to-text system. The videos were uploaded from 2009-2019. 

\subsection{Classroom Datasets Used For Testing}
\subsubsection{NCTE Dataset}
The NCTE dataset \cite{demszky2022ncte} is a collection of video and audio recordings of 2128 4th and 5th-grade mathematics classrooms. Out of these recordings, only 17 are transcribed, highlighting the difficulty of large-scale human transcription. These recordings amount to 12.8 hours of speech. We set aside 4 classroom recordings for a 2.9-hour test/validation set, and use the rest for training a benchmark model that represents the performance using existing classroom data.
\subsubsection{M-Powering Teachers (MPT) Dataset}

The MPT dataset is a collection of 6 classroom recordings that took place in 2023. These classrooms were distributed between California, Ohio, and Washington D.C., and range from the 5th to the 8th grade. The total duration of this dataset is 3 hours. We reserve this dataset entirely for testing.

\section{Methodology}
\label{sec: method}
\subsection{Creating The Clean Classroom Speech Base}
To create the clean version of SimClass, we combine the adult corpora from OCW and Khan Academy with tracks from the MyST dataset. To create variety, for each combination, the file can either start with child speech followed by adult speech, or vice versa. In either case, we allow for a random overlap between 0.5 seconds and 1 second for 20\% of the files. Since we had more files for MyST than OCW and Khan Academy, there are some tracks that were just student talk. 

The purpose of combining tracks from adult and child datasets is to simulate conversations between the teacher and students. Overlap simulates when one person cuts the other off. Although the linguistic content between both phrases does not match, this does not affect models without an embedded language model like Wav2vec. In any case, in a dynamic environment like classrooms, an abrupt change of subject within a topic (e.g. STEM) is expected and can make the model more robust.

The total duration of the SimClass dataset is 391 hours, making it the largest classroom speech corpus and the only publicly available one. As for the train/test/validation partition, we partitioned each constituent dataset before the combination. We follow the splits created by the authors of the MyST dataset which ensures that no speaker exists in two splits. For the YouTube datasets, we partitioned them by channels to create a roughly 80/10/10 partition across both OCW and Khan Academy. As a result, we ended up with 313 hours of training data, 37 hours of development data, and 41 hours of test data.

\subsection{Simulating Classroom Noises in Unity Game Engine}

A game engine is a software framework used by developers to create video games, interactive simulations as well as Extended Reality (XR) applications. While visual components are often the primary focus, audio fidelity plays a crucial role in enhancing immersion, particularly in XR \cite{xr-audio1,Reyes-Lecuona2022}. Consequently, significant research and development efforts have been dedicated to improving audio rendering quality, with a particular emphasis on spatial audio techniques \cite{broderick2018importance, gilberto2024virtual}.

Physical and acoustic simulations have long been an alternative use of game engines \cite{nieminen2021unity, 9793395, 9540074}. One major advantage of using game engines for acoustic simulation is their built-in spatialization and environmental acoustics modeling, which significantly reduces the complexity of designing realistic soundscapes. Many essential acoustic properties, such as occlusion, reverberation, and sound propagation, are integrated into the engine, allowing researchers to focus primarily on structuring the environment visually rather than implementing low-level acoustic physics. While parameter tuning is required to refine the accuracy of sound interactions, it remains far simpler than developing a complete acoustic simulation from scratch.

To further enhance realism, game engines support advanced spatial audio plugins such as Steam Audio \cite{SteamAudio}, which provides real-time physics-based sound propagation and binaural spatialization. Steam Audio leverages the game’s geometric environment to model effects such as diffraction, occlusion, and dynamic reverberation, offering a more accurate representation of how sound interacts with the virtual space. A key feature of Steam Audio is its support for ‘acoustic material’. Using acoustic material, one can design the acoustic properties of any surface, such as absorption, transmission, and scattering, which are then applied to the geometric 3D shapes of different objects. For example, one can design the acoustic properties of wood which can then be applied to  objects like chairs and desks, and each object will influence sound propagation based on both its geometry and assigned acoustic properties.

We thus create our classroom environment inside the Unity Game engines. We acoustically design all objects and surfaces in the classroom, including chairs, desks, doors, whiteboards, windows, ceilings, and carpets. We place 20 spatial directive audio sources spread around each playing track from the non-transcribed portion of the MyST Tutor dataset. Spatial and directive audio sources are used to simulate realistic classroom acoustics, where sound is not only emitted but also interacts with the environment based on its directionality and position. Each source is assigned a specific orientation and emission pattern, allowing for a more accurate representation of how speech propagates within a simulated space. These overlapped audio sources represent the main noise source in our classroom environment, which is children's babble noise. 

In addition to babble noise, we also play random spatial chair noises from random locations in the classroom environment at random intervals. We also play ambient non-directive non-spatial noises in the background that represent outside noises in the classroom like playground noises. These noise tracks were sourced from YouTube.

For audio capture, we have an "audio listener'' or a microphone that moves around the room in order to capture noise from different angles. Through this setup, we capture 50 hours of realistic, physically simulated classroom noise. The simulation was designed on an M3 Macbook Pro but compiled to run on lightweight Linux machines. 
\section{Experiments}
\label{sec: exp}
\subsection{ASR Experiments}
To validate our data, we run ablation ASR experiments comparing the effectiveness of training a Wav2vec2.0-based model on our data and noise versus off-the-shelf datasets. We considered two off-the-shelf datasets, Librispeech \cite{panayotov2015librispeech} and TEDLIUM \cite{hernandez2018ted}. Librispeech is a popular dataset and is a standard choice for developing ASR models. TEDLIUM is a collection of transcribed TED talks which makes it a good point of comparison for instructional speech. For the SimClass dataset, we consider different configurations. We first train on  the clean dataset to measure its efficacy compared to other off-the-shelf datasets. We also create two noisy versions of SimClass, one with our simulated classroom noise, and one using adult babble noise from the FreeSound noise corpus\cite{font2013freesound}. For each noisy dataset, we mix the noises at different signal-to-noise ratios (SNRs), uniformly sampled between -5 and 15 dB with 5dB increments. 

As for the pre-trained Wav2vec model, we consider two models. The first is W2V-Robust \cite{hsu2021robust} as it was pre-trained on noisy English speech. We also consider W2V-Classroom\cite{attia2024cpt}, a Wav2vec model adapted particularly for classroom speech.

\begin{table}[]
\centering
\caption{ASR WER performance with Wav2vec models fine-tuned with different training data tested on the NCTE and MPT test sets. Freesound refers to the Adult babble noises from the Freesoud noise corpus.}
\vspace{-15pt}
\label{tab:asr-NCTE}
\resizebox{\columnwidth}{!}{%
\begin{tabular}{cccc}
 \multicolumn{1}{c}{}& \multicolumn{1}{c}{} & \multicolumn{2}{c}{\textbf{Testset}} \\
\hline
\textbf{Pretrained Model} & \textbf{Training Set} & \textbf{NCTE} & \textbf{MPT} \\\hline
\textbf{W2V-Robust} & \textbf{Librispeech} & 40.64 & 47.59 \\
\textbf{W2V-Robust} & \textbf{TEDLIUM} & 55.82 & 59.63 \\
\textbf{W2V-Robust} & \textbf{SimClass-Clean} & 38.59 & 39.98 \\
\textbf{W2V-Robust} & \textbf{SimClass-Noisy (FreeSound)}& 35.11 & 36.61 \\
\textbf{W2V-Robust} & \textbf{SimClass-Noisy} & 32.88 & 35.74 \\\hline
\textbf{W2V-Classroom} & \textbf{SimClass-Noisy} & 26.29 & 31.72 \\
\textbf{W2V-Classroom} & \textbf{SimClass-Noisy + NCTE}& \textbf{19.63} & \textbf{28.52} \\
\textbf{W2V-Classroom} & \textbf{NCTE} & 21.12 & 32.44\\\hline
\end{tabular}%
}
\vspace{    -15pt}
\end{table}

Table \ref{tab:asr-NCTE} shows the performance of the models in Word Error Rate (WER). Looking at the first three rows of the table, we see that training on SimClass-Clean outperforms Librispeech and TEDLIUM in both the NCTE and the MPT datasets, with significant improvement in the MPT dataset. This shows that our proposed clean speech dataset is a better match for classroom speech than off-the-shelf datasets. Adding noise to the training data improves the performance across the board, however, our proposed simulated classroom noise significantly outperforms adult babble noise from FreeSound. 

Changing the pre-trained model from W2V-Robust to W2V-Classroom significantly improves the performance with both test sets, which is in line with previous findings \cite{attia2024cpt}. Compared with training directly on a limited amount of classroom speech as seen in the last row of the table (NCTE), training on SimClass-Noisy outperforms training on real classroom data with the MPT test set. However, training on NCTE training data outperforms our proposed training data with the NCTE test set due to a better match between the training and test data distributions. Combining SimClass Noisy with NCTE training data leads to the best configuration with both test sets.
\begin{figure}[h!]
    \centering
\vspace{-10pt}
\includegraphics[width=\columnwidth]{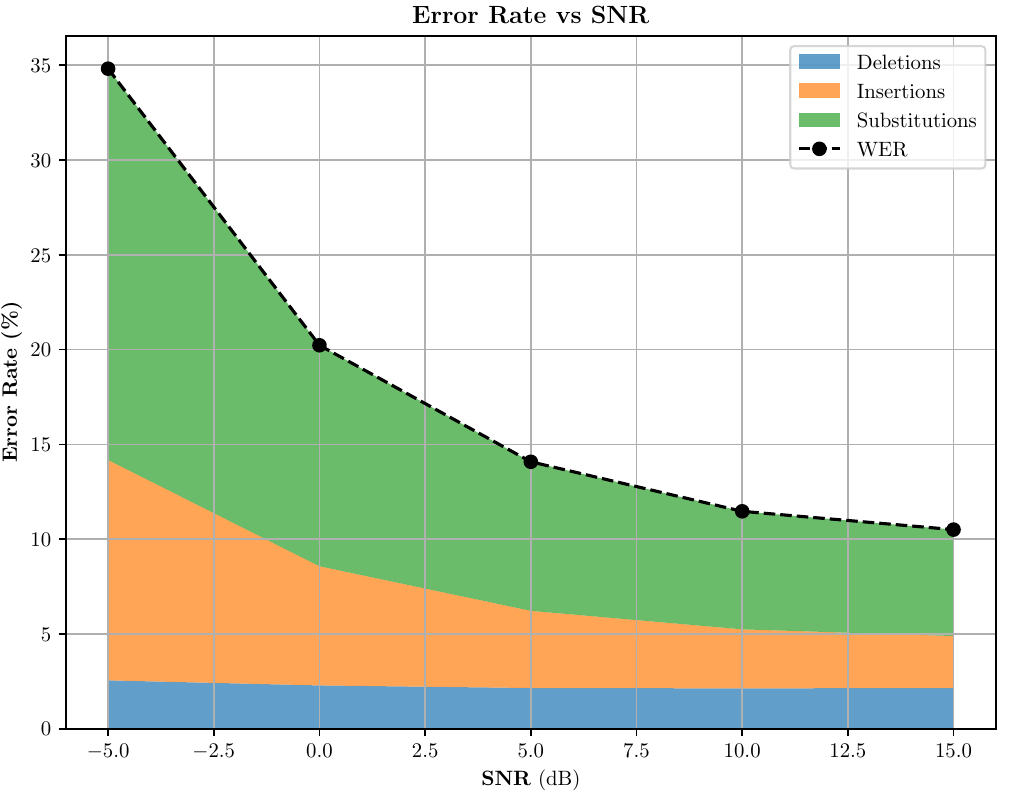}
\vspace{-15pt}
    \caption{Breakdown of word error rate (WER) for W2V-Classroom fine-tuned on a combination of SimClass noisy and NCTE training sets. The model is evaluated on the SimClass test set mixed with noise at varying SNR levels. The WER is decomposed into insertions, deletions, and substitutions, which sum up to the total WER at each SNR.}\vspace{-5pt}
    \label{fig:simclass-snr}
\end{figure}
Figure \ref{fig:simclass-snr} shows the performance of the model trained on SimClass Noisy and NCTE training sets, tested on different versions of the SimClass tests at different SNRs. While more analysis of these results is needed, such data was not possible with actual classroom recordings as there is no way to vary the noise level for the same classroom recording under the same conditions.  It is interesting to note that as noise levels increase, substitution errors are the most affected, followed by insertion errors, while deletion errors remain largely unchanged across different noise levels.

\subsection{Speech Enhancement Experiments}
The separation between noise and clean speech in the SimClass dataset opens the door for new tasks in classroom speech AI not previously possible with real classroom recordings, such as speech enhancement. To showcase that, we fine-tune the StoRM diffusion-based speech enhancement model \cite{lemercier2023storm}. We fine-tune the StoRM model, initially trained on the Voicebank/DEMAND\cite{valentini2016investigating} dataset, using the SimClass-noisy train set mixed at different SNRs. We chose this pre-trained model because Voicebank/DEMAND is a well-established speech enhancement dataset that includes babble noise in its training set, making it a suitable match for our SimClass speech enhancement task. 

Given the slow inference of the diffusion model, we evaluate only 100 randomly sampled files from the SimClass test set.We ensure diversity in the selection and confirm stability across multiple runs. Table \ref{tab:enhancement} shows key metrics from both the fine-tuned and off-the-shelf models on the sampled SimClass test set mixed with simulated classroom noise at SNRs from -5 to 15 dB at 5 dB intervals. We evaluate speech quality using Perceptual Evaluation of Speech Quality (PESQ) \cite{rix2001perceptual}, intelligibility with Extended Short-Term Objective Intelligibility (ESTOI) \cite{jensen2016algorithm}, and overall signal fidelity with Scale-Invariant Signal-to-Distortion Ratio (SI-SDR) \cite{le2019sdr}.
\begin{table}[h!]
\centering
\vspace{-5pt}
\caption{Key speech enhancement metrics of the StoRM diffusion speech enhancement model both off-the-shelf and fine-tuned on the SimClass training data, when tested on the SimClass test data at different SNRs.}
\label{tab:enhancement}
\resizebox{0.65\columnwidth}{!}{%
\begin{tabular}{cccc}

\hline
\multicolumn{4}{c}{\textbf{StoRM Finetuned on SimClass}} \\\hline
\textbf{SNR} & \textbf{PESQ} & \textbf{ESTOI} & \textbf{SI-SDI} \\\hline
\textbf{-5} & \textbf{1.388}  &\textbf{ 0.596}  &\textbf{8.093} \\
\textbf{0} & \textbf{1.678}  & \textbf{0.710}   &\textbf{11.131} \\
\textbf{5} & \textbf{2.034}  & \textbf{0.787 }   &\textbf{14.157} \\
\textbf{10} & \textbf{2.453}  & \textbf{0.842}  &\textbf{16.988}  \\
\textbf{15} &\textbf{2.861} & \textbf{0.882}   &\textbf{19.795} \\\hline
\multicolumn{4}{c}{\textbf{StoRM Off-The-Shelf}} \\\hline
\textbf{SNR} & \textbf{PESQ} & \textbf{ESTOI} & \textbf{SI-SDI} \\\hline
\textbf{-5} &  1.085  & 0.308  & -2.815 \\
\textbf{0} & 1.167 &  0.501 &  4.288\\
\textbf{5} & 1.376 &  0.644 &  9.132\\
\textbf{10} &  1.605 &  0.743  & 12.827 \\
\textbf{15} &   1.923 &  0.805 &  15.530\\\hline
\end{tabular}%
}
\end{table}
Table \ref{tab:enhancement} demonstrates that fine-tuning StoRM with SimClass data leads to consistent and significant improvements across all metrics and SNR levels.  While the performance at low SNRs still has low integrability and perceptual quality, our results highlight the value of SimClass as a realistic and challenging dataset for classroom speech enhancement. 

Furthermore, we test our enhanced speech signals with the ASR system trained on the SimClass noisy training data and the NCTE dataset. Table \ref{tab:enhanced-ASR} shows the results. While speech enhancement improves the results, the improvement is marginal at SNRs above -5. It is a documented phenomenon that speech enhancement introduces artifacts that negatively impact ASR systems \cite{lemercier2023storm}. To that end, we also mix the noisy signal with the enhanced signal, a technique previously shown to improve ASR robustness by mitigating enhancement artifacts \cite{fujimoto2019one}.

\begin{table}[]
\centering
\caption{ASR WER performance with the W2V-Classroom model fine-tuned on SimClass Noisy and NCTE training sets, evaluated on a subset of the SimClass test set. The test set is either mixed with noise at different SNR levels, enhanced using our speech enhancement model, or a mixture of enhanced and noisy signals.}
\vspace{-10pt}
\label{tab:enhanced-ASR}
\resizebox{0.65\columnwidth}{!}{%
\begin{tabular}{cccc}
\hline
\textbf{SNR} & \textbf{Noisy} & \textbf{Enhanced} & \textbf{Mixed} \\\hline
\textbf{-5} & 33.01 & 29.63 & \textbf{22.35} \\
\textbf{0} & 17.17 & 17.06 & \textbf{13.38} \\
\textbf{5} & 12.65 & 12.35 & \textbf{10.63} \\
\textbf{10} & 9.85 & 9.19 & \textbf{8.82} \\
\textbf{15} & 9.49 & 8.79 & \textbf{8.24} \\\hline
\textbf{Clean} & \multicolumn{3}{c}{7.68} \\\hline

\end{tabular}%
}
\end{table}

\section{On the Effect of Diversity of Noise on Generalizability}
In our experiments with the FreeSound noise corpus, we only used the noise files with adult babble noise. In this section, we experiment with using all noise categories in the FreeSound corpus, including Car, AC, Metro, Cafe, Traffic as well as Adult Babble. 

\begin{table}[]
\centering
\caption{ASR WER performance with Wav2vec models fine-tuned with different noisy training data tested on the NCTE and MPT test sets. Freesound Babble Only refers to the Adult babble noises from the Freesoud noise corpus, while FreeSound - All refers to the entire FreeSound training corpus.}
\vspace{-10pt}
\label{tab:asr-diverse}
\resizebox{\columnwidth}{!}{%
\begin{tabular}{ccc}
\multicolumn{1}{c}{} & \multicolumn{2}{c}{\textbf{Testset}} \\
\hline
\textbf{Training Set} & \textbf{NCTE} & \textbf{MPT} \\\hline
\textbf{SimClass-Noisy (FreeSound Babble Only)}& 35.11 & 36.61 \\
\textbf{SimClass-Noisy  (FreeSound - All)} & \textbf{32.63} & \textbf{35.58} \\
\textbf{SimClass-Noisy} & 32.88 & 35.74 \\
\hline
\end{tabular}%
}\vspace{-10pt}
\end{table}

Table \ref{tab:asr-diverse} shows the results from three models, each trained on the SimClass dataset mixed with different noise corpora. We can see that including all kinds of noise, even those unrelated to the classroom domain, such as ''Car" and ''Traffic", notably improves the performance compared to just training on ''Adult Babble", and matches the performance of our classroom simulated noise. This highlights the effect of the diversity of the additive noise on the generalizability and noise robustness of the ASR model. This also identifies an important room for improvement for the SimClass simulated noise in future editions. In the meantime, we encourage researchers to experiment with mixing the simulated noise with other kinds of noise, as that might improve the noise robustness of the ASR model.   
\section{Conclusion and Future Research Directions}
In this paper we present the SimClass classroom speech dataset, the largest and first public classroom speech dataset. We also propose a novel and versatile method of simulating noises utilizing game engines. Our ablation studies show our proposed data's efficacy in two key tasks, ASR and speech enhancement. In ASR, our training data outperforms popular off-the-shelf datasets like Librispeech and TEDLIUM. When paired with our proposed simulated noise, our model training on real classroom datasets shows that our data realistically simulates classroom data. In speech enhancement, SimClass shows promise as the first classroom dataset suitable for speech enhancement training, although further research is required in this area.

We have identified key areas of improvement for our data for future versions. Currently, student and teacher speech are added together without taking into consideration their relative positions to each other and the microphone. Additionally, room acoustics are not taken into consideration for this process. We aim to incorporate Room Impulse Responses in the combination of the audio tracks for the clean version of the dataset. Additionally, we aim to introduce greater variety in the noise simulation, focusing on new noise sources in addition to speech and generating noises from different classroom environments.
\bibliographystyle{IEEEtran}
\bibliography{mybib}

\end{document}